\documentclass[aip, amsmath,amssymb]{revtex4-1}

\usepackage{xcolor}
\usepackage{graphicx}% Include figure files
\usepackage{dcolumn}% Align table columns on decimal point
\usepackage{bm}% bold math
\usepackage[utf8]{inputenc}
\usepackage[T1]{fontenc}
\usepackage[english]{babel}
\usepackage{mathptmx}
\usepackage{amsmath,amssymb}
\begin{document}
	\title[Complementarity between Brillouin signature and Scholte wave for controlled elasticity in fibrillated collagen medium for culture cell.]{Complementarity between Brillouin signature and Scholte wave for controlled elasticity in fibrillated collagen medium for culture cell.}
	% Force line breaks with \\
	\author{A. Hamraoui}
	\email{ahmed.hamraoui@u-paris.fr}
	\affiliation{Université Paris Cité, CNRS UMR 8003, Saints-Pères Paris Institute for the Neurosciences,   45 rue des Saints-Pères, 75006 Paris, France}%
	\affiliation{Sorbonne Université, CNRS UMR7574, Laboratoire de Chimie de la Matière Condensée de Paris, 4 place Jussieu, 75005 Paris, France}%
	\author{O. Sénépart}%
	\affiliation{Université Paris Cité, CNRS UMR 8003, Saints-Pères Paris Institute for the Neurosciences,   45 rue des Saints-Pères, 75006 Paris, France}%
	\affiliation{Sorbonne Université, CNRS UMR7574, Laboratoire de Chimie de la Matière Condensée de Paris, 4 place Jussieu, 75005 Paris, France}%
	% \email{Second.Author@institution.edu.}
	\author{L. Belliard}
	\affiliation{Sorbonne Université, CNRS UMR7588, Institut des Nanosciences de Paris, 4 place Jussieu, 75005 Paris, France}%

	\date{\today}% It is always \today, today,
	
	\begin{abstract}
		Modulating extracellular matrix (ECM) elasticity with fibrillar collagen offers great potential for regenerative medicine, drug discovery, and disease modeling by replicating \textit{in-vivo} mechanical signals. This enhances understanding of cellular responses and fosters therapeutic innovation. However, precise ECM elasticity measurements are still lacking. This study couples time-resolved Brillouin spectroscopy and pulsed laser-induced Scholtes wave generation.  We measure how collagen fibrillation affects sound velocity and refractive index. These insights are advancing tissue engineering and cellular biomechanics. 
	\end{abstract}
	
	\maketitle
	
	\section{Introduction :}
	Collagen, making up 30\% of the body's proteins, is prevalent in bones, muscles, blood, and skin, where it constitutes 75\%. It is crucial for skin elasticity and serves as an ideal matrix for joints, tendons, and ligaments. Understanding collagen fibrillation kinetics is essential in tissue engineering and cell culture, where optimal conditions for cell growth and function are paramount. Manipulating the mechanical properties of the extracellular matrix (ECM) is crucial as ECM elasticity significantly influences cellular processes like adhesion, migration, proliferation, and differentiation. Fibrillar collagen, with its adjustable mechanical properties, stands out due to its ability to influence cellular behavior and tissue development. By controlling collagen concentration, cross-linking density, and fiber alignment, researchers can tailor ECM stiffness and elasticity to mimic various tissue microenvironments, enhancing the relevance of in-vitro models\cite{cortese_influence_2014,keshmiri_brillouin_2024,elsayad_mapping_2016}. This modulation has significant implications for regenerative medicine, drug discovery, and disease modeling, aiding in understanding cellular responses to different conditions.
	
	Despite its potential, quantitative ECM elasticity measurements remain incomplete. Brillouin spectroscopy, which is particularly well suited for investigations of semitransparent systems, has already been applied to the elastic characterization in numerous biological studies.\cite{cortese_influence_2014,keshmiri_brillouin_2024,elsayad_mapping_2016}.  In this letter, we report on alternative experiments on the thermoelastic excitation of acoustic waves induced by a pulsed laser in a collagen solution. Simultaneous detection of longitudinal waves and Scholte waves at the collagen/solid interface enables us to quantitatively measure the effect of the fibrillation process on sound velocity and refractive index versus the fibrillation time, providing insights into the relationship between collagen fibrillation and ECM mechanical properties.

	\section{Material and method }
	Collagen fibrillation was performed directly on the supports, which were 1 x 1 cm silicon wafers, for variable durations ranging from 30 minutes to 72 hours.  The collagen utilized was concentrated to \(20 \,mg/mL\), specifically SYMATESE collagen, which contained \(17 \,mM\) acetic acid. For each sample, a mixture of \(0.387 \,mL\) collagen, \(0.018 \,mL\) sodium hydroxide, and \(0.044 \,mL\) PBS was prepared. The mixture was stirred to ensure uniform distribution of sodium hydroxide and PBS within the collagen. Subsequently, the dish was resealed and placed in the culture room’s incubator for the required fibrillation duration.
	
	Time resolved pump and probe experiments have been realized using mode-locked Ti:sapphire (MAI-TAI Spectra) laser source operated at 800 nm with a pulse duration below 100 fs at the laser output and a repetition rate of 79.4 MHz. Synchronous interferometric detection is performed by modulating the pump beam at 1.8 MHz using an acousto-optic modulator. A  Michelson scheme allowed to measure the perpendicular surface displacement.  The pump and probe are first superimposed on the acoustic transducer to study longitudinal acoustic waves. To obtain information on surface or interface waves, the probe beam is swept around the pump excitation using a 4f configuration. In both configurations we performed a two-color pump-probe experiment by doubling the frequency of the probe or pump beam. 
	As the fibrillation process starts at the free surface of the collagen, we used a sufficiently thin metallic transducer, deposited on a sapphire or glass substrate, placed in contact with this free surface. The titanium or aluminium layer is thick enough to generate an acoustic burst and thin enough to allow the probe beam to follow the propagation of the acoustic pulse in the first few micrometers of the biomaterial layer. 
	\section{Results and discussion}
\begin{figure}[ht]
		\centering
		\includegraphics[width=0.7\linewidth]{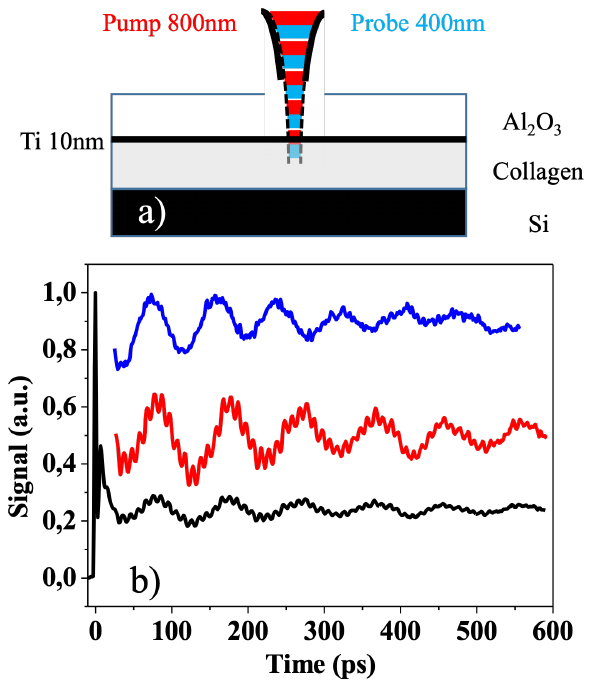}
		\caption{a) Experimental geometry. b) Time resolved measurement. Bottom: Raw data with collagen layer with a fibrillation time of 24h. Middle: Signal showing the Brillouin oscillation without the background. Top: Brillouin oscillation with a fibrillation time of 48h.}
		\label{fig:figure1}
	\end{figure}

	First let us consider longitudinal acoustic wave propagating perpendicular to the surface, using the experimental geometry described in figure \ref{fig:figure1}a. In this case, the advantage of sapphire, compared with a glass substrate for example, lies in the fact that the Brillouin signature of this transparent compound is in a much higher frequency range, around \(100\,GHz\), than that expected in collagen gel. In Figure \ref{fig:figure1}b, reflectivity variations are measured with variable pump and probe delay. After subtraction of the photothermal background, a double Brillouin oscillation can be clearly observed. The high-frequency signature comes from to the acoustic wave transmitted into the sapphire, while the low-frequency component corresponds to the collagen layer signature.  In this case, the transducer should be pressed sufficiently on the collagen layer to ensure good acoustic transfer.
	
\begin{figure}[ht]
		\centering
		\includegraphics[width=0.7\linewidth]{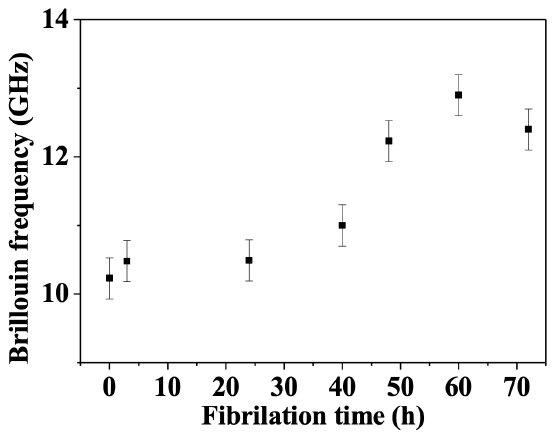}
		\caption{Variation of Brillouin frequency in collagen 20mg/ml versus the fibrillation time.}
		\label{fig:figure2}
	\end{figure}

	The lower frequency component evolves as a function of fibrillation time. The evolution of this Brillouin frequency versus the fibrillation time is shown in Figure \ref{fig:figure2}. Over a 72-hour period, the frequency changes by almost 26\%. This frequency is linked to the layer's elastic properties by the following formula: 
	\begin{equation}
		F_{\text{Brillouin }}=\dfrac{2 n}{\lambda} V
	\end{equation}
	where $\lambda$ is the probe wavelength, here $400 \,nm$, V the longitudinal velocity and n the refractive index.	
	
\begin{figure}
		\centering
		\includegraphics[width=0.7\linewidth]{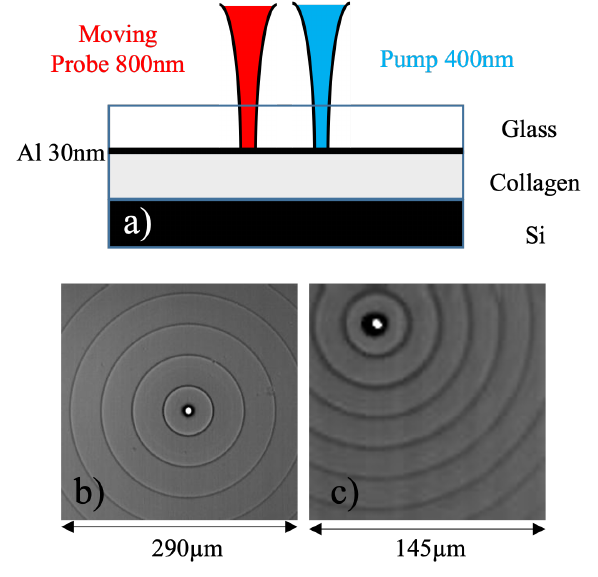}
		\caption{a) Experimental geometry. The Al transducer thickness is chosen to have enough acoustic emission in the plane and enough probe transmission to detect interface waves. b) Rayleigh wave mapping at the free surface c) Scholte wave mapping at the interface Al-collagen.}
		\label{fig:figure3}
	\end{figure}

\begin{figure}
		\centering
		\includegraphics[width=0.7\linewidth]{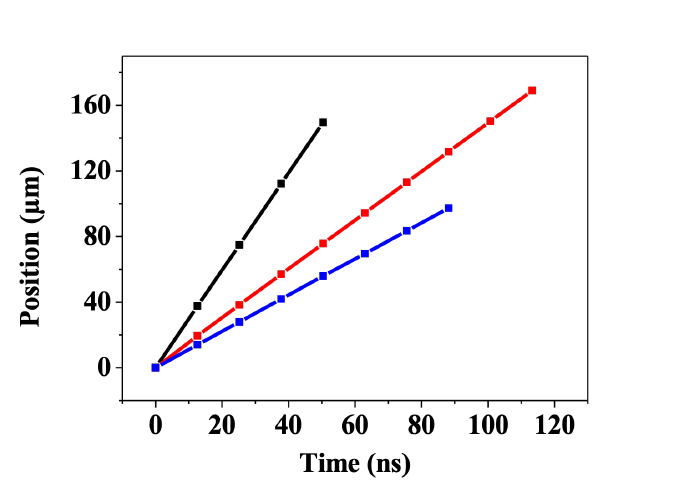}
		\caption{Position of the wave front at different time with wave velocity deduced. In black: data for a Rayleigh wave. In red: data's for a Scholte wave in collagen. In blue Scholte wave in ethanol.}
		\label{fig:figure4}
	\end{figure}

	Consequently, knowledge of the refractive index is mandatory to determine longitudinal velocity. This parameter varies considerably in the literature and is also likely to change with fibrillation time. 
	To overcome this limitation, we used an alternative geometry to directly measure the wave velocity in the layer, independently of the optical properties. While our previous focus was on longitudinal volume waves, we are now concentrating on waves that may exist at the interface between the metal transducer and the collagen gel.\\		
	Following this, the probe beam is scanned around the pump epicenter to be sensitive to in-plane propagation in the collagen layer. The experimental setup is briefly illustrated in figure \ref{fig:figure3}a.  The mapping of surface displacements when the collagen layer is simply replaced by air is shown in figure \ref{fig:figure3}b.
	A set of concentric circles corresponding to the Rayleigh waves emitted by the various pump pulses can be observed. \cite{PhysRevB.73.125403}.
	The isotropic nature within the plane is associated with the isotropy of the glass substrate used here. Figure \ref{fig:figure3}c shows a similar situation when the aluminum transducer is in contact with non-fibrillated collagen. In this case, we note the presence of similar features which clearly exhibited a much lower propagation velocity. 
	By pointing out the position of these rings and knowing the repetition rate of our laser, \(12.59\,ns\) in this case, we obtain figure \ref{fig:figure4}, enabling us to extract the propagation velocity of these different waves.
	For an air layer, a velocity of \(3020\,m/s\) is obtained, which is in perfect agreement with a Rayleigh wave propagating in glass substrate \cite{mante_complete_2008}. Here, the very thin aluminum layer has no influence on this propagation due to the low frequency of the surface acoustic wave involved. \\
	In the case of ethanol or collagen, we obtain much lower velocities: 1118 m/s and 1507m/s respectively, indicating that we are dealing with a different type of wave.
	In this hard solid / fluid configuration\cite{glorieux_character_2001,gusev_theory_1996}, where the transducer has a much higher longitudinal velocity than the liquid, it is well known that the Rayleigh wave is transformed into a strongly attenuated leaky Rayleigh wave, additionally, a Scholte wave appears at the interface, capable of propagating over long distances. The velocity \(v_{s}\) of this wave is given by the following equation:		
	\begin{equation}
		\begin{aligned}
			& \left(2-\left(\frac{v_s}{v_{t 1}}\right)^2\right)^2-4\left(1-\left(\frac{v_s}{v_{t 1}}\right)^2\right)^{\frac{1}{2}}\left(1-\left(\frac{v_s}{v_{l 1}}\right)^2\right)^{\frac{1}{2}}+ \\
			& \frac{\rho_2}{\rho_1}\left(1-\left(\frac{v_s}{v_{l 1}}\right)^2\right)^{\frac{1}{2}}\left(1-\left(\frac{v_s}{v_{l 2}}\right)^2\right)^{-\frac{1}{2}}\left(\frac{v_s}{v_{t 1}}\right)^4=0 
		\end{aligned} 
	\end{equation}

	Where $v_{\ell}$ and $v_t$ are the longitudinal and transverse velocity  respectively, $\rho$ the density, indices 1 and 2 refer to solid and liquid respectively.
	\\For glass-ethanol interface the Scholte wave velocity is expected to be very close to the longitudinal wave velocity of ethanol equal to 1150m/s, which is in good agreement with the experimental speed deduced from figure \ref{fig:figure4}, demonstrating that the concentric rings depicted in figure \ref{fig:figure3}b may be associated to such Scholte waves.  Scholte waves induces by laser excitation has already been demonstrated in lower frequency range\cite{desmet_laser-induced_1996}.
	Conversely, by introducing the Scholte velocity measured in a collagen solution into equation (2), we can deduce a longitudinal velocity of \(1512\, m/s\) for non-fibrillated collagen. Then, considering this deduced longitudinal velocity, and the Brillouin frequency measured in figure \ref{fig:figure2}, we obtain a value for refractive index at \(400\,nm\) equal to \(1.35\). This new approach, which enables the refractive index to be determined using purely acoustic measurements, sheds new light on issues relating to the elasticity of living cells, where this parameter has hitherto been neglected due to a lack of quantitative measurements \cite{liu_remote_2019,perez-cota_new_2019,viel_picosecond_2019,liu_labelfree_2019,danworaphong_three-dimensional_2015,viel_picosecond_2019}. 
	
\begin{figure}
		\centering
		\includegraphics[width=0.7\linewidth]{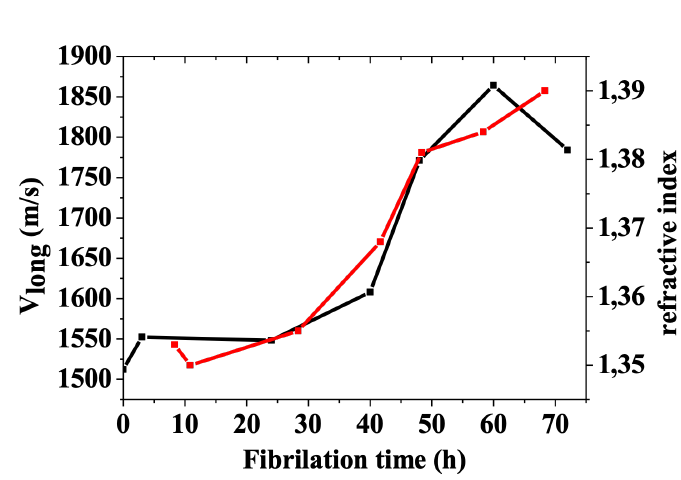}
		\caption{Deduced longitudinal velocity ( in black) and optical index ( in red) versus the fibrillation time.}
		\label{fig:figure5}
	\end{figure}

	By reproducing the same acoustic mappings for fibrillated solutions, we can extract the evolution of longitudinal velocity and refractive index as a function of fibrillation time, without having to rely on the knowledge of additional parameters. These evolutions are shown in figure \ref{fig:figure5}. Consequently, the fluctuation in Brillouin frequency observed in figure \ref{fig:figure2} was in fact essentially linked to the variation in collagen elasticity, with the optical index varying by only around 3\%. Now, if the density is assumed to be roughly constant equal to 1, then the bulk modulus giving by $B= \rho V^{2}$ may change of 40\% varying to 2.5GPa up to 3.5GPa.   
	It should nevertheless be kept in mind that the treatment proposed here is based on the assumption that the longitudinal velocities in the plane and perpendicular to the layer are equal, potential anisotropies in the propagation properties are neglected.

	\section{Conclusion}
	We have developed a new technique to determine the longitudinal velocities and refractive index in materials whose elasticity can be modified on demand, which is particularly relevant for biological systems such as biomaterials. This method has shown remarkable efficiency in studying cell growth, allowing for variations of up to 40\% in the bulk elastic modulus. The application of this technique opens new perspectives for the analysis and engineering of biomaterials, providing precise and adaptable means to optimize mechanical properties according to the specific needs of biological studies. However, a major challenge remains: the precise determination of the material's density, a crucial step for the complete characterization of its mechanical properties. Continuing research in this direction could not only enhance our understanding of biomaterials but also lead to significant advancements in their practical applications.
	
	\bibliographystyle{unsrt}
	%	\nocite{*}
	\bibliography{FibrillationCollagene}% Produces the bibliography via BibTeX.
	
\end{document}